\documentclass[prb,aps,twocolumn,superscriptaddress,showpacs,dvips]{revtex4}
\usepackage{feynmp}
\usepackage{amssymb}
\usepackage{epsfig}
\usepackage{graphicx}
\usepackage{subfigure}
\usepackage{hyperref}
\usepackage{bbm}

\begin{document}

\title{Suppression of superconductivity at nematic critical point in underdoped cuprates}

\author{Guo-Zhu Liu}
\affiliation{Max Planck Institut f$\ddot{u}$r Physik komplexer
Systeme, D-01187 Dresden, Germany} \affiliation{Department of Modern
Physics, University of Science and Technology of China, Hefei, Anhui
230026, P. R. China}
\author{Jing-Rong Wang}
\affiliation{Department of Modern Physics, University of Science and
Technology of China, Hefei, Anhui 230026, P. R. China}
\author{Jing Wang}
\affiliation{Department of Modern Physics, University of Science and
Technology of China, Hefei, Anhui 230026, P. R. China}

\begin{abstract}
A nematic quantum critical point is anticipated to exist in the
superconducting dome of some high-temperature superconductors. The
nematic order competes with the superconducting order and hence
reduces the superconducting condensate at $T = 0$. Moreover, the
critical fluctuations of nematic order can excite more nodal
quasiparticles out of the condensate. We address these two effects
within an effective field theory and show that superfluid density
$\rho^s(T)$ and superconducting temperature $T_c$ are both
suppressed strongly by the critical fluctuations. The strong
suppression of superconductivity provides a possible way to
determine the nematic quantum critical point.
\end{abstract}

\pacs{71.10.Hf, 73.43.Nq, 74.20.De}

\maketitle


\section{Introduction}

The strong electron correlation in high-temperature superconductors
(HTSC) is able to drive an electronic nematic phase \cite{Kivelson,
Kivelson03, Fradkin, Vojta}, which preserves translational symmetry
but breaks rotational symmetry. In the past decade, there have been
a number of experimental signatures pointing to the presence of
nematic ordering transition in some HTSCs \cite{Fradkin, Vojta,
Ando, Hinkov, Daou, Lawler}. On the basis of these experiments, a
zero-temperature nematic quantum critical point (QCP) is supposed to
exist at certain doping concentration $x_c$ in the superconducting
(SC) dome \cite{Fradkin, Vojta, VojtaSachdev, Kim, Huh, Xu, Fritz,
Lawler2, Liu}. Generally, the nematic order has two impacts on the
SC state. First, it competes with the SC order \cite{Fradkin, Vojta,
Kivelson03}. Second, the nematic order parameter couples to the
gapless nodal quasiparticles (QPs), which are believed to be the
most important fermionic excitations in unconventional
superconductors with $d_{x^2 - y^2}$ energy gap. The latter coupling
is singular at the nematic QCP $x_c$, and has stimulated
considerable theoretical efforts \cite{VojtaSachdev, Kim, Huh, Xu,
Fritz, Lawler2, Liu}. A recent renormalization group analysis
\cite{Huh} showed that it leads to a novel fixed point at which the
ratio between gap velocity $v_{\Delta}$ and Fermi velocity $v_F$ of
nodal QPs flows to zero, $v_{\Delta}/v_F \rightarrow 0$.

\begin{figure}[ht]
\centering
  \vspace{-0.3cm}
   \includegraphics[width=3.3in]{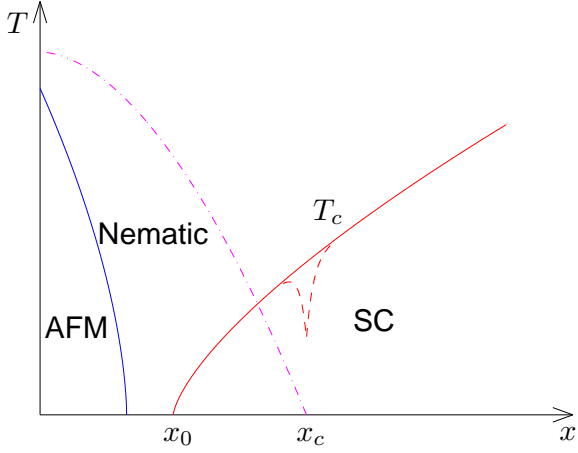}
  \vspace{-0.3cm}
\caption{Schematic phase diagram of HTSC. The suppression of $T_c$
at nematic QCP $x_c$ is represented by the dashed
line.}\label{Fig:1}
\end{figure}

Although a zero-temperature nematic QCP is expected to exist
somewhere in the SC dome \cite{Fradkin, VojtaSachdev, Kim, Huh, Xu,
Fritz, Lawler2, Liu}, shown schematically in Fig. (\ref{Fig:1}), its precise
position, and even its very existence, has not been unambiguously
confirmed by experiments so far. It is therefore always interesting
to seek evidence which can help convincingly confirm or disconfirm
the existence of such point. In this paper, we study the superfluid
density $\rho^s(T)$ and the SC temperature $T_c$ at the supposed
nematic QCP $x_c$. If $\rho^s(T)$ and $T_c$ exhibit sharply distinct
behaviors at $x_c$, then the nematic QCP may be detected by
measuring these quantities.

HTSCs are known to be doped Mott insulators, so their superfluid
density is much smaller than that of conventional metal
superconductors. At $T = 0$, the superfluid density in underdoping
region depends \cite{Orenstein00, Orenstein90} linearly on doping
$x$ as $\rho^{s}(0) = x/a^2$, where $a$ is the lattice spacing. At
finite $T$, certain amount of nodal QPs are thermally excited out of
the SC condensate. Lee and Wen argued that these normal nodal QPs
can efficiently deplete the superfluid density \cite{Lee97}.
Formally, the superfluid density contains two terms, $\rho^{s}(T) =
\rho^{s}(0) - \rho^{n}(T)$, where $\rho^{n}(T)$ is the normal QPs
density. Setting $\rho^{s}(T_c) = 0$ allows for an estimate of the
critical temperature $T_c$. Employing a phenomenological approach,
Lee and Wen \cite{Lee97} obtained $T_c \propto \rho^s(0) \propto
\frac{v_\Delta}{v_F} x$, reproducing the Uemura plot \cite{Uemura}.

Once a nematic ordering transition occurs at $x_c$, the superfluid
density and $T_c$ will be substantially changed. As $v_{\Delta}/v_F
\rightarrow 0$ due to the critical nematic fluctuations, it seems
that $T_c \rightarrow 0$, i.e., superconductivity would be
completely suppressed at $x_c$. This argument is certainly
oversimplified since the above expression of $T_c$ is obtained in
the non-interacting limit. However, this qualitative analysis does
indicate the importance of the critical nematic fluctuations, and
indeed motivates us to perform a quantitative computation of the
renormalized $\rho^{s}(T)$ and $T_c$ after taking into account the
nematic fluctuations.

The nematic order affects $\rho^{s}(T)$ in two ways. On the one
hand, since the nematic order competes with the SC order, it reduces
$\rho^{s}(0)$. This reduction can be examined by studying the
competitive interaction between nematic and SC order parameters. On
the other, the critical nematic fluctuations can excite more nodal
QPs out of the condensate, compared with the case without nematic
order. As a consequence, $\rho^{n}(T)$ is enhanced and the
superfluid density is further suppressed. We shall access this
effect by generalizing the phenomenological approach proposed in
Ref.~\cite{Lee97}. The velocity anisotropy plays an essential role
in these considerations. After explicit calculations, we find that
superfluid density $\rho^s(T)$ and $T_c$ are both significantly
reduced due to critical nematic fluctuations, indicating a strong
suppression of superconductivity at nematic QCP $x_c$ (see Fig.
(\ref{Fig:1})).

The rest of the paper is organized as follows. In Sec.II, we address
the competitive interaction between SC and nematic order parameters
and calculate zero-$T$ superfluid density. In Sec.III, we calculate
the density of normal QPs after taking into account fermion velocity
renormalization due to critical nematic fluctuation. Based on these
calculations, we predict a strong suppression of superconductivity
at nematic QCP. In Sec.IV, we present a brief summary of our
results, and also discuss the possible experimental determination of
the nematic QCP.

\section{Competing orders and zero-temperature superfluid density}

We first consider the renormalized zero-$T$ superfluid density at
nematic QCP. Based on phenomenological grounds, we write down a free
energy of two competing orders,
\begin{eqnarray}
F &=& F_{\psi} + F_{\phi} + F_{\psi\phi} \nonumber \\
&=& \frac{1}{2m}(\nabla \psi)^2 - \alpha \psi^2 +
\frac{\beta}{2}\psi^4 + F_{\phi} + \gamma \psi^2 \phi^2,\label{Eq_free-energy}
\end{eqnarray}
where $\psi$ and $\phi$ are the SC and nematic order parameters,
respectively. The strength of the competitive interaction between
$\psi$ and $\phi$ is represented by a positive constant $\gamma$.
Such type of effective model has been adopted to describe competing
orders in various superconductors \cite{Demler, Zaanen}.

In the absence of nematic order, the mean value of $\psi$ is $|\psi|
= \sqrt{\alpha/\beta}$. To be consistent with experiments, the
parameters $\alpha$ and $\beta$ must be properly chosen such that
$4|\psi|^2 = 4\alpha/\beta = \rho^{s}(0) = x/a^2$. In the presence
of nematic order, $|\psi|$ will be renormalized by the $\gamma
\psi^2 \phi^2$ term. The quantum fluctuation of nematic order $\phi$
is very strong and actually singular at nematic QCP $x_c$, so $\phi$
should be regarded as a quantum-mechanical field. However, we can
consider $\psi$ as a classical field and ignore its quantum
fluctuations, provided $x_c$ is not close to the SC QCP $x_0$.

The free energy of nematic order, $F_{\phi}$, should be specified
now. Analogous to the free energy of SC order $\psi$, $F_{\phi}$
contains a quadratic term $\phi^2$ and a quartic term $\phi^4$.
However, the additional coupling between nematic order and nodal QPs
introduces an extra term. The action describing this coupling is
given by \cite{VojtaSachdev, Kim, Huh}
\begin{eqnarray}
S_{\Psi} &=& \int \frac{d^3k}{(2\pi)^3}
[\Psi^{\dagger}_{1i}(-i\omega +
v_{F}k_{1}\tau^{z} + v_{\Delta}k_{2}\tau^x)\Psi_{1i} \nonumber \\
&& + \Psi^{\dagger}_{2i}(-i\omega
+ v_{F}k_{2}\tau^{z} + v_{\Delta}k_{1}\tau^{x})\Psi_{2i}], \\
S_{\Psi\phi} &=& \int d^2xd\tau [\lambda_0
\phi(\Psi^{\dagger}_{1i}\tau^{x}\Psi_{1i} +
\Psi^{\dagger}_{2i}\tau^{x}\Psi_{2i})],
\end{eqnarray}
where $\tau^{x,y,z}$ are Pauli matrices and the flavor index $i$
sums up $1$ to $N$. $\Psi_{1}$ represents nodal QPs excited from
$(\frac{\pi}{2},\frac{\pi}{2})$ and
$(-\frac{\pi}{2},-\frac{\pi}{2})$ points, and $\Psi_{2}$ the other
two. The effective action of $\phi$ has the form \cite{Huh}
\begin{eqnarray}
S_{\phi} = \int \frac{d^3q}{(2\pi)^3}\left[\frac{1}{2}\left(q^2 + r
+ \Pi(q)\right)\phi^2 + \frac{u}{4}\phi^{4}\right],\label{Eq_effective_phi}
\end{eqnarray}
which is obtained by integrating out $N$-flavor nodal QPs. Here, $r$
is the tuning parameter for the nematic ordering transition, with $r
= 0$ defining the nematic QCP $x_c$. The polarization function
$\Pi(q)$ comes from nodal QPs, and, to the leading order of
$1/N$-expansion, is defined as
\begin{eqnarray}
\Pi(\epsilon,\mathbf{q}) = N\int\frac{d\omega
d^{2}\mathbf{k}}{(2\pi)^3}
\mathrm{Tr}[\tau^{x}G_{0}(\omega,\mathbf{k})\tau^{x}
G_{0}(\omega+\epsilon,\mathbf{k+q})], \nonumber \\
\end{eqnarray}
where
\begin{eqnarray}
G_0(\omega,\mathbf{k}) = \frac{1}{-i\omega + v_{F}k_1 \tau^{z} +
v_{\Delta}k_2 \tau^{x}}\label{Eq_G_0}
\end{eqnarray}
is the free propagator for nodal QPs $\Psi_1$ (free propagator for
nodal QPs $\Psi_2$ can be similarly written down). The polarization
function $\Pi(\epsilon,\mathbf{q})$ has already been calculated
previously \cite{Huh, Liu}, and is known to have the form
\cite{note}
\begin{eqnarray}
\Pi(\epsilon,\mathbf{q}) = \frac{N}{16v_F
v_\Delta}\left[\frac{\epsilon^2 + v_F^2 q_1^2}{(\epsilon^2 + v_F^2
q_1^2 + v_{\Delta}^2 q_2^2)^{1/2}} + (q_1 \leftrightarrow
q_2)\right]. \nonumber \\
\end{eqnarray}
Note there is no direct interaction between $\psi$ and nodal QPs
$\Psi$. Indeed, $\Psi$ are excited on top of a SC order $\psi$.
Moreover, they have sharp peak and long lifetime in the SC dome in
the absence of competing orders \cite{Orenstein00}, so their
coupling to $\psi$ must be quite weak.

Starting from Eq. (\ref{Eq_free-energy}) and Eq. (\ref{Eq_effective_phi}),
we can compute the correction to
zero-$T$ superfluid density due to the competition between SC and
nematic orders. To this end, we need to minimize the effective
potential ofSC order, $V[\psi]$. The bare potential for $\psi$ is
simply $V_0[\psi] = - \alpha \psi^2 + \frac{\beta}{2}\psi^4$. It
receives an additional term $V_1[\psi]$ due to the SC-nematic
competition. This additional term will be calculated using the
methods presented in Refs.~ \cite{Halperin, Zaanen}. The
corresponding partition function is
\begin{eqnarray}
Z[\psi(\mathbf{r})] = \int \mathcal{D}\phi(\mathbf{r},\tau)
\exp\left(-\frac{\mathcal{F_{\psi}}}{T} - S_{\phi} -
S_{\psi\phi}\right)
\end{eqnarray}
where $\mathcal{F}_{\psi}=\int d^{2}\mathbf{r}F_{\psi}$. The
saddle-point equation for $\psi$ reads
\begin{eqnarray}
\frac{\delta\ln Z[\psi(\mathbf{r})]}{\delta\psi(\mathbf{r})} = 0,
\end{eqnarray}
which gives rise to
\begin{eqnarray}
\left[-\alpha+\beta\psi^2(\mathbf{r}) -
\frac{1}{m}\mathbf{\nabla}^{2} + \gamma
f[\psi]\right]\psi(\mathbf{r}) = 0.
\end{eqnarray}
Here, $f[\psi]$ is the expectation value of $\phi^2$. At the
one-loop level, it has the form
\begin{eqnarray}
f[\psi] \equiv \langle \phi^2 \rangle = \int \frac{d^3 q}{(2\pi)^3}
\frac{1}{q^2 + \Pi(q) + \gamma\psi^2},
\end{eqnarray}
where $\Pi(q)$ and $\gamma\psi^2$ represent the contributions due to
nodal QPs and SC order, respectively. Although the $q^2$ term
appearing in the denominator of $f[\psi]$ is much smaller than
$\Pi(q)$ at low energy, it cannot be simply neglected since
$f[\psi]$ would be divergent without such term. It is now
straightforward to get
\begin{eqnarray}
\frac{\delta V_1[\psi]}{\delta \psi} = 2\gamma\psi f[\psi],
\end{eqnarray}
which then leads to a renormalized potential
\begin{eqnarray}
V[\psi] = - \alpha \psi^2 + \frac{\beta}{2}\psi^4 + V_1[\psi].
\end{eqnarray}
To calculate the renormalized zero-$T$ superfluid density,
$\rho_{\mathrm{R}}^s(0)$, one needs to minimize the effective
potential $V[\psi]$ by taking
\begin{eqnarray}
\frac{\delta V[\psi]}{\delta \psi} = 0.
\end{eqnarray}
The renormalized $|\psi|^2$ and therefore $\rho_{\mathrm{R}}^s(0)$
can be obtained from the solution of the following equation
\begin{eqnarray}
- \alpha \psi + \beta\psi^3 + \gamma\psi f[\psi] = 0.\label{Eq_solution}
\end{eqnarray}

In order to see the role of nodal QPs, we first ignore the fermion
contribution in $f[\psi]$ by assuming $\Pi(q) = 0$. In this case,
the integration over $q$ in $f[\psi]$ can be exactly performed,
yielding
\begin{eqnarray}
f[\psi] = \frac{1}{4\pi}(\Lambda-\sqrt{\gamma}\psi).
\end{eqnarray}
The renormalized potential becomes
\begin{eqnarray}
V[\psi] = V_0[\psi] + \frac{\gamma \Lambda}{2\pi}\psi^2 -
\frac{\gamma\sqrt{\gamma}}{6\pi}\psi^3.
\end{eqnarray}
The cubic term induced by nematic order turns the SC transition to
first order \cite{Halperin, Zaanen}, with critical point being
$\alpha_c = \frac{\gamma \Lambda}{4\pi} - \frac{\gamma^3}{64\pi^2
\beta}$. However, since the gapless nodal QPs are present even at
the lowest energy, the polarization function $\Pi(q)$ should be
included in the effective action $S_{\phi}$. After including the
polarization $\Pi(q)$ into $f[\psi]$, integration over $q$ can not
be carried out analytically, and numerical method will be used.
After numerically solving Eq. (\ref{Eq_solution}), we found a
critical value $\gamma_c$ for the competitive interaction between SC
and nematic orders. The SC transition remains continuous if $\gamma
< \gamma_c$, and is driven to first order when $\gamma > \gamma_c$.

\begin{figure}[ht] \centering
   \includegraphics[width=3.2in]{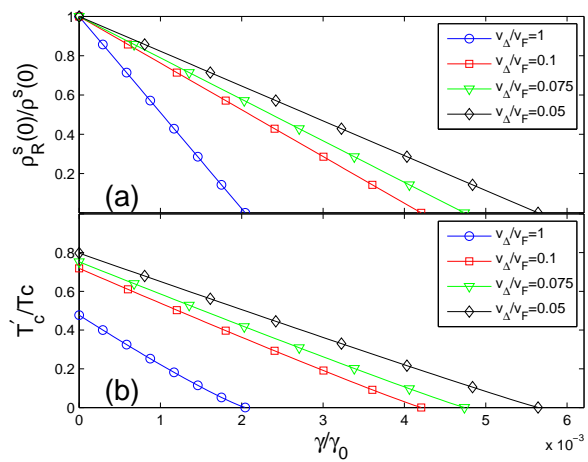}
   \vspace{-0.2cm}
\caption{(a) $\rho_{\mathrm{R}}^{s}(0)/\rho^{s}(0)$ for four
different values of bare ratio $v_\Delta/v_F$; (b)
$T_c^{\prime}/T_c$. The open circles \emph{etc.} are guides to
eyes.} \label{Fig:Rho0RG}
\end{figure}

We now discuss the effects of the competing nematic order on the
zero-$T$ superfluid density. When $\gamma$ is zero or very small,
the nematic order does not change $|\psi|^2$, which is expected and
trivial. As $\gamma$ increases continuously, $|\psi|^2$ and
$\rho_{\mathrm{R}}^s(0)$ both decrease rapidly. To estimate this
effect more quantitatively, we assume an ultraviolet cutoff $\Lambda
= 10\mathrm{eV}$, and choose $\alpha = 2.5 \times
10^{-3}\mathrm{eV}$. Moreover, we consider a representative bare
velocity ratio $v_\Delta/v_F \approx 0.075$, which is an appropriate
value for YBa$_2$Cu$_3$O$_{6+\delta}$ \cite{Chiao}. We further
choose $\rho^{s}(0)/m = 4\alpha/m\beta = 10^{-2}\mathrm{eV}$, which
corresponds to $T_c \approx 20$K. It is now useful to introduce a
parameter given by $\gamma_0 = \beta/\xi_0^2 \alpha$ with $\xi_0^2 =
1/2m\alpha$, and define a dimensionless coupling constant
$\gamma/\gamma_0$. Fig. \ref{Fig:Rho0RG}(a) shows that
$\rho_{\mathrm{R}}^s(0)$ is strongly suppressed by the nematic order
and is completely destroyed when $\gamma$ is large enough. It is
also obvious that bare velocity ratio plays a vital role: a larger
anisotropy causes smaller drop of $\rho^s(0)$. This property
provides further evidence that nodal QPs can not be simply ignored.

\section{Velocity renormalization and suppression of
superconductivity}

We then turn to calculate the density of normal nodal QPs,
$\rho^n(T)$. We first briefly outline the phenomenological approach
of Lee and Wen \cite{Lee97} and then generalize it to the case with
critical nematic fluctuation. In the scenario of Lee and Wen, a
crucial problem is how to assess the roles played by nodal and
antinodal QPs in the destruction of SC condensate. As revealed
clearly by extensive experiments \cite{Shen}, a pseudogap phase
exists above $T_c$. The pseudogap turns out to have the same
$d$-wave symmetry as the SC gap \cite{Shen}, which implies the
antinodal QPs remain gapped as temperature increases across $T_c$.
It was argued \cite{Lee97} that the SC state is primarily destroyed
by the thermal proliferation of the low lying, gapless nodal QPs,
and the antinodal QPs are only spectators. Lee and Wen further
assume that the nodal QPs do not carry superflow \cite{Lee97}, so
the fermion spectrum is shifted by the vector potential to
$E(\mathbf{k},\mathbf{A}) = E(\mathbf{k}) +
\frac{e}{c}\mathbf{v}_{k}\cdot \mathbf{A}$, where the QPs energy is
linearized as $E(\mathbf{k} ) = \sqrt{v_{F}^{2}k_{1}^{2} +
v_{\Delta}^{2}k_{2}^{2}}$ near nodes $(\frac{\pi}{2},
\frac{\pi}{2})$ and $\mathbf{v}_{k}$ is the normal state velocity.
The electric current is $j_{\mu} = \frac{e}{c m}
\rho_{\mu\nu}^{s}\mathbf{A}_{\nu}$, where the superfluid tensor
$\rho_{\mu\nu}^{s}$ can be written as $\rho_{\mu\nu}^{s}(T) =
\rho^{s}(0)\delta_{\mu\nu} - \rho_{\mu\nu}^{n}(T)$. The zero-$T$
superfluid density is $\rho^{s}(0) = x/a^2$, and the normal QPs
density is derived from the free energy \cite{Lee97}
\begin{eqnarray}
F(\mathbf{A},T) = -T \sum_{\mathbf{k},\sigma} \ln\left(1 +
e^{-E(\mathbf{k},\mathbf{A})/T}\right)
\end{eqnarray}
by the formula $\frac{1}{m}\rho_{\mu\nu}^{n}(T) =
-2\sum_{\mathbf{k}}\frac{d E}{d A_{\mu}} \frac{d E}{d
A_{\nu}}\frac{\partial n_{F}}{\partial E}$. In the non-interacting
case, the fermion velocities are constants, so one gets
\begin{eqnarray}
\frac{1}{m}\rho^s(T) = \frac{x}{m a^2} -
\frac{2\ln2}{\pi}\frac{v_F}{v_{\Delta}} T,
\end{eqnarray}
which exhibits a linear temperature dependence, in agreement with
experiments \cite{Hardy}. At $T = 0$, $\rho^s(T)$ takes its maximum
value, $\rho^s(0)$. As $T$ is increasing from zero, $\rho^s(T)$
decreases rapidly due to thermally excited nodal QPs and eventually
vanishes as $T \rightarrow T_c$. By taking $\rho^s(T_c) = 0$, it is
easy to obtain
\begin{eqnarray}
T_c \propto \rho^s(0) \propto \frac{v_{\Delta}}{v_F}\frac{x}{m a^2}.
\end{eqnarray}
This linear doping dependence of $T_c$ is well consistent with the
Uemura plot. The same results were later reproduced by means of
Green's function method \cite{Paramekanti}. The Fermi liquid (FL)
corrections to these results were also investigated \cite{Millis,
Paramekanti}. An important fact is that both $\rho^s(T)$ and $T_c$
contain the velocity ratio $v_{\Delta}/v_F$, so these two quantities
are expected to be significantly affected by the critical nematic
fluctuation which can cause a nontrivial renormalization of the
velocity ratio.

The above approach of computing $T_c$ is applicable when the nodal
QPs are well defined. At the nematic QCP $x_c$, the critical
fluctuation of nematic order couples strongly to gapless nodal QPs
and may lead to breakdown of FL behavior. As a consequence, the
nodal QPs might no longer be well defined QPs. In a strict sense,
the validity of the simple $d$-wave BCS theory \cite{Lee97} and its
FL-interaction generalizations \cite{Millis, Paramekanti} are both
doubtful. In order to proceed, here we assume that the basic
approach of Ref.~\cite{Lee97} is still valid after the free energy
of QPs receives a singular correction due to nematic fluctuation.

\begin{figure}[ht]
\centering
\includegraphics[width=3in]{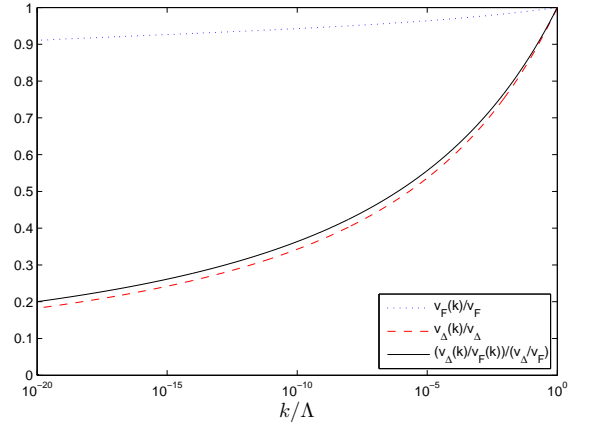}
   \vspace{-0.2cm}
\caption{Flow of $v_F$, $v_\Delta$, and $v_\Delta/v_F$ with momentum
$k$ for bare ratio $v_\Delta/v_F = 0.075$. As $k \rightarrow 0$,
both velocities are reduced down to zero. However, $v_\Delta$
decreases much faster than $v_F$, so velocity ratio $v_\Delta/v_F
\rightarrow 0$.} \label{Fig:v}
\end{figure}

The free fermion propagator shown in Eq. (\ref{Eq_G_0}) is renormalized by
nematic order to
\begin{eqnarray}
G(\omega,\mathbf{k}) = \frac{1}{G_0^{-1}(\omega,\mathbf{k}) -
\Sigma(\omega,\mathbf{k})},
\end{eqnarray}
where the self-energy is
\begin{eqnarray}
\Sigma(\omega,\mathbf{k}) = \int \frac{d\epsilon
d^2\mathbf{q}}{(2\pi)^3}
G_0(\omega+\epsilon,\mathbf{k}+\mathbf{q})\frac{1}{q^2 + \Pi(q)}.
\end{eqnarray}
The pole in $G(\omega,\mathbf{k})$ determines the renormalized
energy $\widetilde{E}(\omega,\mathbf{k}) = E(\omega,\mathbf{k}) +
\delta E(\omega,\mathbf{k})$, where $E(\omega,\mathbf{k})$ is given
by pole of free propagator $G_0(\omega,\mathbf{k})$ and $\delta
E(\omega,\mathbf{k})$ represents the interaction correction to
energy. To describe this renormalization procedure, it is convenient
to expand the self-energy formally as
\begin{eqnarray}
\Sigma(\omega,\mathbf{k}) = -i\omega\Sigma_0 + \Sigma_1 v_F k_1
\tau^z + \Sigma_2 v_\Delta k_2 \tau^x,
\end{eqnarray}
which exhibits unusual logarithmic behaviors since $\Sigma_{1,2}(k)$
contains such a term as $\ln(\Lambda/k)$. We combine this
self-energy with $G_0^{-1}(k)$, and find that the bare velocities
acquire strong $k$-dependence, $v_{F,\Delta} \rightarrow
v_{F,\Delta}(k)$. $v_{F,\Delta}(k)$ are very complicated functions
of $k$, and thus can not be written down analytically. After
straightforward numerical computation, we show the $k$-dependence of
fermion velocities and their ratio in Fig. (\ref{Fig:v}), for the
representative bare value $v_{\Delta}/v_F = 0.075$. It is clear that
both $v_{F}(k)$ and $v_{\Delta}(k)$ vanish as $k \rightarrow 0$.
Nevertheless, $v_{F}(k)$ approaches zero much more slowly than
$v_{\Delta}(k)$ does. Therefore, the velocity ratio vanishes at the
lowest energy, $v_\Delta/v_F \rightarrow 0$, which recovers the
extreme velocity anisotropy \cite{Huh}. Such velocity
renormalization causes breakdown of FL behavior \cite{Xu, Kim97} and
can faithfully characterize the non-FL nature of QPs at $x_c$.

\begin{figure}[ht]
\centering
\includegraphics[width=3.2in]{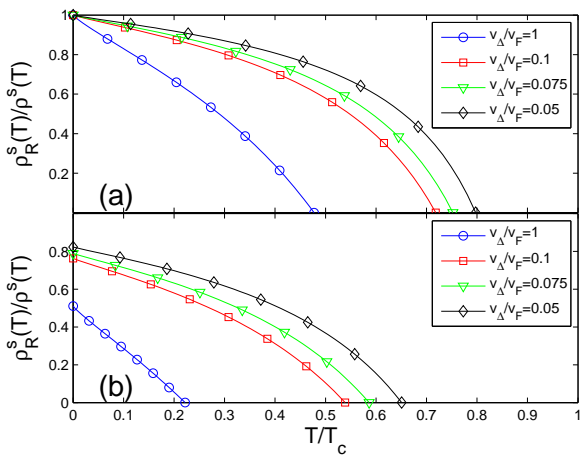}
   \vspace{-0.2cm}
\caption{Superfluid density at finite temperature,
$\rho_{\mathrm{R}}^{s}(T)$. (a): $\gamma = 0$; (b): $\gamma/\gamma_0
= 10^{-3}$.} \label{Fig:RhoT}
\end{figure}

Although the nodal QPs are strongly damped by the nematic
fluctuations and thus no longer well-defined at $x_c$, it is still
reasonable to assume that they do not carry superflow \cite{Lee97},
which allows us to write $\widetilde{E}(\mathbf{k},\mathbf{A}) =
\widetilde{E}(\mathbf{k}) +
\frac{e}{c}\mathbf{v}_{k}\cdot\mathbf{A}$, where the normal velocity
$\mathbf{v}_{k}$ becomes strongly $k$-dependent. Now the new free
energy is
\begin{eqnarray}
\widetilde{F}(\mathbf{A},T) = -T \sum_{\mathbf{k},\sigma} \ln(1 +
e^{-\widetilde{E}(\mathbf{k},\mathbf{A})/T}),
\end{eqnarray}
from which we obtain a renormalized superfluid density
\begin{eqnarray}
\rho_{\mathrm{R}}^s(T) &=& \rho_{\mathrm{R}}^s(0) -
\rho_{\mathrm{R}}^{n}(T), \\
\frac{\rho_{\mathrm{R}}^{n}(T)}{m} &=& \frac{4}{T}\int\frac{d^2
\mathbf{k} }{(2\pi)^{2}} \frac{v_{F}^{2}(k)e^{\sqrt{v_{F}^{2}
(k)k_{1}^{2} +v_{\Delta}^{2}(k)k_{2}^{2}}/T}}{\left(1 +
e^{\sqrt{v_{F}^{2}(k)k_{1}^{2} +
v_{\Delta}^{2}(k)k_{2}^{2}}/T}\right)^{2}}.
\end{eqnarray}
At nematic QCP, $T_c$ is changed from its original value to a
renormalized value $T_c^{\prime}$. When $T = T_c^{\prime}$, the
renormalized superfluid density vanishes,
$\rho_{\mathrm{R}}^s(T_c^{\prime}) = 0$, so that
\begin{eqnarray}
\rho_{\mathrm{R}}^s(0) = \rho_{\mathrm{R}}^{n}(T'_{c}),
\end{eqnarray}
which builds a relationship between $T'_{c}$ and $T_{c}$. The ratios
of $\rho_{\mathrm{R}}^{s}(T)/\rho^{s}(T)$ and $T_c^{\prime}/T_c$ can
be numerically calculated according to the above three equations. To
include the influence of interaction corrections, the velocities
$v_{F,\Delta}$ appearing in $\rho_{\mathrm{R}}^s(0)$ are also
replaced by $v_{F,\Delta}(k)$. The numerical results of
$T_c^{\prime}/T_c$ are shown in Fig. \ref{Fig:Rho0RG}(b), and those
of $\rho_{\mathrm{R}}^{s}(T)/\rho^{s}(T)$ given in Fig.
(\ref{Fig:RhoT}). The suppression of superconductivity is prominent
even when the competitive interaction between SC and nematic orders
is quite weak, and is further enhanced by this interaction. As
aforementioned, the SC phase transition at $T_c^{\prime}$ remains
continuous as long as $\gamma$ is small.

Clearly, the extent to which $\rho^s(T)$ and $T_c$ are suppressed by
the nematic order depends on a number of parameters, such as
$\alpha$, $\gamma$, and $v_\Delta/v_F$. It is helpful to make more
quantitative analysis. The magnitudes of $\alpha$ as well as
$\Lambda$ have already been chosen. The bare velocity ratio is still
fixed at $v_\Delta/v_F = 0.075$. Fig. \ref{Fig:Rho0RG}(b) shows that $T_c$
is roughly suppressed by $25\%$ when $\gamma = 0$, and by $50\%$ when
$\gamma/\gamma_0 \approx 1.5 \times 10^{-3}$. The drop of $\rho^{s}(T)$ can
be analyzed in a similar way. An important implication of Fig. (\ref{Fig:RhoT})
is that $\rho^{s}(T)$ is more strongly reduced
at higher temperatures. Experimentally, the suppression of $T_c$ can
be tested by measuring resistivity and Meissner effect, and the
suppression of superfluid density may be probed by measuring London
penetration length $\lambda_L$, preferably using microwave
techniques \cite{Hardy, Basov} based on the relationship $\rho^s(T)
\propto \lambda_L^{-2}(T)$.

The strong suppression effects rely on the divergence of correlation
length $\xi$ of the nematic field $\phi$ at the QCP $x_c$. As one
moves away from the point $x_c$, the strong suppression of
superfluid density and $T_c$ will be rapidly diminished. In fact,
once correlation length $\xi$ becomes finite, the interaction
between nematic fluctuation and nodal QPs is no longer singular.
Consequently, the self-energy of nodal QPs, $\Sigma(k)$, does not
display logarithmic behavior, which indicates that the fermion
velocities only receive unimportant renormalizations. In this case,
the density of normal QPs is not significantly modified by the
competing nematic order, since the velocity ratio does not depart
far from its bare value. The influence of nemtic order on
$\rho^s(0)$ is also weakened. For $x > x_c$, the fluctuation of
$\phi$ is rather weak and does not change $\rho^s(0)$ much if
$\gamma$ is not very large. For $x_0 < x < x_c$, the fluctuation of
$\phi$ is gapped and also leads to much smaller change of
$\rho^s(0)$ compared with what happens at nematic QCP $x_c$.
Therefore, the $T_c$ curve generally follows the Uemura plot in the
whole underdoped region, but deviates from the linear behavior in
the close vicinity of $x_c$. Since the suppression of
superconductivity is most pronounced at $x_c$, it may help to find
the position of the predicted nematic QCP if such a point really
exists.

\section{Summary and discussion}

In summary, we have considered the effects of nematic fluctuation on
superfluid density and critical temperature $T_c$ in the vicinity of
nematic QCP in the contexts of some HTSCs. On one hand, the critical
fluctuation of nematic order parameter reduces zero-$T$ superfluid
density as a result of ordering competition. On the other hand, it
couples strongly to the gapless nodal QPs and excites more normal
QPs out of the SC condensate by triggering an extreme fermion
velocity anisotropy. Both of these two effects combine to
significantly suppress superconductivity. Therefore, we have
predicted a dip shape reduction of $T_c$ at nematic QCP $x_c$, which
is schematically shown in Fig. (\ref{Fig:1}).

The dip shape of $T_c$ reminds us of $1/8$ anomaly \cite{Kivelson03,
Fradkin, Vojta, Moodenbaugh, Tranquada, Valla, Sato}, which refers
to an anomalous suppression of superconductivity at doping $x =
0.125$. A sudden drop of $\rho^s(0)$ was also observed
\cite{Panagopoulos} in La$_{2-x}$Sr$_{x}$CuO$_4$ (LSCO) at $x =
0.125$. It appears that $\rho^s(T)$ and $T_c$ near the nematic QCP
bear some similarity to the basic features of these experiments.
However, we refrain from identifying the $1/8$ doping as the
anticipated nematic QCP for several reasons: i) $1/8$ anomaly is
usually attributed to the formation of static stripe order
\cite{Tranquada, Kivelson03}; ii) it is not clear why nematic QCP
exists precisely at $x = 0.125$, not elsewhere; iii) low-$T$ dc
thermal conductivity $\kappa/T$ was predicted to be enhanced at
nematic QCP \cite{Fritz}, whereas an early transport measurement
found a drop of $\kappa/T$ in LSCO at $x = 0.125$ \cite{Takeya}.

Recently the critical nematic fluctuations are shown to induce
unusual behaviors in several quantities, such as QPs spectral
function \cite{Kim}, specific heat \cite{Xu}, nuclear relaxation
rate \cite{Xu}, thermal conductivity \cite{Fritz}, and QP
interference \cite{Lawler2}. Normally, a single observation alone is
not able to uniquely fix the nematic QCP. Should all or most of
these predictions, including the suppression of superconductivity,
be observed, the nematic QCP might be determined. However, the
absence of a sharp nematic QCP does not necessarily mean the absence
of nematic phase, since the sharp QCP may be rounded by disorders
and become a crossover \cite{Kim}. In case this smearing occurs,
many of the anomalous critical behaviors are weakened, or even
destroyed, and the nematic phase most possibly shows its existence
in the $\omega \ll T$ regime.

In the calculations presented in this paper, we have assumed that
the SC order parameter $\psi$ is a classical field and neglected its
quantum fluctuation. This assumption is expected to be a good one if
the nematic QCP $x_c$ is deep inside the SC dome, namely $x_c$ is
not close to the SC critical point $x_0$. When $x_c$ is close to
$x_0$, the quantum fluctuation of SC order $\psi$ may become very
important and a fully quantum-mechanical treatment will be
necessary.

\section*{Acknowledgement}

G.Z.L. acknowledges support by the National Natural Science
Foundation of China under grant No. 11074234 and the Visitors
Program of MPIPKS at Dresden.

\end{document}